# Exchange and Distribution of Multipartite Entanglement via Atomic Coherence

Xihua Yang[1], Bolin Xue[1], Junxiang Zhang[2], Shiyao Zhu[3]

[1]*Department of Physics, Shanghai University, Shanghai 200444, China*

[2]*State Key Laboratory of Quantum Optics and Quantum Optics Devices, Institute of Opto-Electronics, Shanxi University, Taiyuan 030006, China*

[3]*Beijing Computational Science Research Center, Beijing 100084, China*

(Dated: Jul. 7, 2014)

We present a convenient and efficient way to exchange and distribute multipartite entangled state via atomic coherence induced by electromagnetically induced transparency (EIT) in an atomic ensemble. By using the EIT-based nondegenerate four-wave mixing processes, the generation, exchange, and distribution of light-light, atom-light, and atom-atom multipartite entanglement can be achieved in a deterministic way with only coherent light fields. This EIT-induced atomic coherence acts as a quantum information processor and has the function of quantum beam splitter, quantum frequency converter, quantum entangler, and quantum repeater, which greatly facilitates the operations in quantum information processing, and holds promising applications in realistic scalable quantum communication and quantum networks.

PACS numbers: 42.50.Gy, 03.67.Bg, 42.50.Dv, 42.65.Lm

Entanglement provides an essential resource for quantum computation, quantum communication, and quantum networks [1-3]. As is well known, light is the best long-distance quantum information carrier, whereas the atomic ensembles provide an attractive medium for storage and manipulation of quantum information. How to conveniently and efficiently realize the generation, distribution, storage, retrieval, and control of light-light, light-matter, as well as matter-matter multipartite entanglement is the basic requirement for realistic quantum information processing. One commonly-used method to generate multiple entangled fields is to combine squeezed fields produced through parametric down-conversion processes in nonlinear optical crystals with linear beam splitters or polarizing beam splitters; however, the created entangled fields are degenerate and suffer from short correlation time. At the nodes of quantum networks, multiple entangled fields with different frequencies are necessary to connect different physical systems. The creation of nondegenerate multipartite entanglement has been investigated by using nonlinear optical processes. Nussenzveig *et al.* demonstrated the production of pump-signal-idler three-color entanglement in the above-threshold optical parametric oscillator (OPO) [4], and a scheme based on this three-color entanglement was proposed to generate scalable multipartite entanglement by using many OPOs operated in a chain configuration [5]; other ways have also been proposed by employing either cascaded nonlinearities or concurrent parametric oscillation [6]. Nevertheless, these schemes are still of limited use due to relatively short correlation time. An alternative attractive avenue is the implementation of the electromagnetically induced transparency [7] (EIT)-based multi-Λ-type systems in an atomic ensemble. By using nondegenerate four-wave mixing (FWM) or Raman scattering processes, the generation of nondegenerate narrow-band multi-entangled fields with long correlation time have been actively studied [8-11]; in addition, the entangled state exchange between an atomic ensemble and light fields [8], as well as between two atomic ensembles [12, 13], have also been realized, which are quite suitable for quantum memory and quantum networks. However, in these EIT-based schemes, the input laser fields for generating entanglement are treated classically and their quantum features are not explored so far.

In quantum communication and quantum networks, the distribution of quantum states over long distance is limited by inevitable photon loss in the transmission channel. To overcome this limitation, Briegel *et al.* [14] introduced the concept of quantum repeater, combining entanglement swapping and quantum memory to extend communication distances. One conventionally-employed strategy for realizing quantum repeater is to use atomic ensembles as quantum memories, combing with linear optical techniques and photon counting to perform all required operations [13]. However, in such protocols, nonclassical input light fields and interferometric stabilization over long distance are required for entanglement swapping; moreover, the entanglement is created in a post-selection way.

Motivated by the experimental observation of generating multi-field correlations and anti-correlations via atomic spin coherence in $^{85}$Rb atomic system [15], here, with the consideration of the quantum properties of the input laser fields, we propose a convenient and efficient scheme to exchange and distribute multipartite entangled state via atomic coherence established by the strong on-resonant coupling and probe fields in the Λ-type EIT configuration. By employing the EIT-based nondegenerate four-wave mixing processes, the generation, exchange, and distribution of light-light, light-atom, as well as atom-atom multipartite entanglement can be conveniently and flexibly achieved in a deterministic way with only coherent light fields. This method would greatly simplify the practical implementation of quantum information processing, and may find potential applications in quantum communication and quantum networks.

The considered model, as shown in Fig. 1a, is based on the experimental configuration in Ref. [15], where the relevant energy levels and applied/generated laser fields form a triple-Λ-type system. Levels $|1\rangle$, $|2\rangle$, and $|3\rangle$ correspond, respectively, to the ground-state hyperfine levels $5S_{1/2}$ (F=3), $5S_{1/2}$ (F=2), and the excited state $5P_{1/2}$ in $D_1$ line of $^{85}$Rb atom with the ground-state hyperfine splitting of 3.036 GHz. The probe field $E_p$ (with frequency $\omega_p$ and Rabi frequency $\Omega_p$) and coupling field E$_c$ (with frequency $\omega_c$ and Rabi frequency $\Omega_c$) are relatively strong and tuned to resonance with the transitions $|2\rangle$-$|3\rangle$ and $|1\rangle$-$|3\rangle$, respectively. By applying a third mixing field E$_{m1}$ (or E$_{m2}$) with frequency $\omega_{m1}$ (or $\omega_{m2}$), off-resonantly coupling levels $|2\rangle$ (or $|1\rangle$) and $|3\rangle$ with detuning $\Delta_1=\omega_{m1}$-$\omega_{32}$ (or $\Delta_2=\omega_{m2}$-$\omega_{31}$), a Stokes field E$_1$ (or an anti-Stokes field E$_2$) can be created through the nondegenerate FWM process. In fact, as shown in Fig. 1b, the produced Stokes field E$_1$ (or anti-Stokes field E$_2$) can be equivalently regarded as scattering the field E$_{m1}$ (or E$_{m2}$) off the atomic coherence $\sigma_{12}$ pre-established by the strong coupling and probe fields in the Λ-type EIT configuration formed by levels $|1\rangle$, $|2\rangle$, and $|3\rangle$. In Ref. [11], we have employed the similar scheme to generate arbitrary number of entangled fields and create quantum entangler via the pre-created atomic coherence, however, the quantum properties of the scattering fields and the entangled feature between the scattering fields and the generated fields have not been investigated. In fact, as shown in Ref. [15], correlations and anti-correlations between the scattering and generated fields have been experimentally observed via atomic coherence. In what follows, by using the Heisenberg-Langevin method with the scattering fields and generated Stokes/anti-Stokes fields treated quantum mechanically, we show how the atomic coherence can act as a quantum information processor and can be employed to realize the exchange and distribution of multipartite entangled state.

By using the equivalent configuration shown in Fig. 1b, we first investigate the entangled feature among the scattering filed E$_{m1}$, Stokes field E$_1$, and atomic coherence $\sigma_{12}$. We assume that the detunings $\Delta_1$ and $\Delta_2$ are sufficiently large so that the coupling between different scattering fields can be neglected. Also, we assume that the coupling and probe fields are substantially strong, so that the atomic coherence is strong enough to ensure that different scattering fields have negligible influence on it. Under these conditions, the Heisenberg-Langevin equations for describing the evolution of the collective atomic operators $\sigma_{12}(z,t)$, $\sigma_{13}(z,t)$, and

$\sigma_{23}(z,t)$ can be written as [16]

$$\dot{\sigma}_{12}(z,t) = -\gamma_0 \sigma_{12} - ig_1 a_1(z,t)\sigma_{23}^+ + ig_2 b_1^+(z,t)\sigma_{13} + F_{12}(z,t), \tag{1}$$

$$\dot{\sigma}_{13}(z,t) = -[\gamma_{13} - i\Delta_1]\sigma_{13} + ig_1 a_1(z,t)(\sigma_{11} - \sigma_{33}) + ig_2 b_1 \sigma_{12} + F_{13}(z,t), \tag{2}$$

$$\dot{\sigma}_{23}(z,t) = -(\gamma_{23} - i\Delta_1)\sigma_{23} + ig_2 b_1(z,t)(\sigma_{22} - \sigma_{33}) + ig_1 a_1 \sigma_{12}^+ + F_{23}(z,t), \tag{3}$$

where $\gamma_{13} = \gamma_{23} = \frac{\gamma_1 + \gamma_2}{2}$ with $\gamma_1$ and $\gamma_2$ being the population decay rates from level 3 to levels 1 and 2, $\gamma_0$ is the coherence decay rate between levels 1 and 2, $a_1$ and $b_1$ are the quantum operators of the Stokes field $E_1$ and scattering field $E_{m1}$, $g_{1(2)} = \mu_{13(23)} \cdot \varepsilon_{1(2)} / \hbar$ is the atom-field coupling constant with $\mu_{13(23)}$ as the dipole moment for the 1-3 (2-3) transition and $\varepsilon_{1(2)} = \sqrt{\hbar \omega_{1(2)} / 2\varepsilon_0 V}$ as the electric field of a single Stokes (scattering) photon with V as the interaction volume with length L and beam radius r, and $F_{ij}(z,t)$ are the collective atomic $\delta$-correlated Langevin noise operators. In the similar analysis described in [16], under the assumption that the uniformly-distributed pencil-shaped atomic sample is optically thin in the transverse direction, the evolution of the operators $a_1$ and $b_1$ can be described by the coupled propagation equations

$$(\frac{\partial}{\partial t} + c\frac{\partial}{\partial z})a_1(z,t) = ig_1 N \sigma_{13}, \tag{4}$$

$$(\frac{\partial}{\partial t} + c\frac{\partial}{\partial z})b_1(z,t) = ig_2 N \sigma_{23}, \tag{5}$$

where N is the total number of atoms in the quantum volume. As done in Ref. [16], we use the perturbation analysis to treat the interaction of the atoms with the fields. In the zeroth-order perturbation expansion, by semiclassically treating the interaction of the atoms with the strong coupling and probe fields in the $\Lambda$-type EIT configuration, we get the steady-state mean values of $\sigma_{11}^{(0)}$, $\sigma_{22}^{(0)}$, $\sigma_{33}^{(0)}$, $\sigma_{12}^{(0)}$, $\sigma_{13}^{(0)}$ and $\sigma_{23}^{(0)}$. By substituting the zeroth-order solution into the Fourier-transformed Heisenberg-Langevin equations for $\sigma_{12}(z,t)$, $\sigma_{13}(z,t)$, and $\sigma_{23}(z,t)$, we can get the first-order solution $\sigma_{12}^{(1)}(z,\omega)$, $\sigma_{13}^{(1)}(z,\omega)$, and $\sigma_{23}^{(1)}(z,\omega)$, which are expressed as

$$\sigma_{12}^{(1)}(z,\omega)=\frac{1}{\gamma_0+i\omega}[-ig_1\sigma_{23}^{+(0)}a_1+ig_2\sigma_{13}^{(0)}b_1^++F_{12}], \qquad (6)$$

$$\sigma_{13}^{(1)}(z,\omega)=\frac{1}{\gamma_{13}+i(\omega-\Delta_1)}[ig_1(\sigma_{11}^{(0)}-\sigma_{33}^{(0)})a_1+ig_2\sigma_{12}^{(0)}b_1+F_{13}], \qquad (7)$$

$$\sigma_{23}^{(1)}(z,\omega)=\frac{1}{\gamma_{23}+i(\omega-\Delta_1)}[ig_2(\sigma_{22}^{(0)}-\sigma_{33}^{(0)})b_1+ig_1\sigma_{12}^{+(0)}a_1+F_{23}], \qquad (8)$$

Note that in the above first-order analysis, we only consider the steady-state mean values of the zeroth-order atomic operators (neglecting the noise operators as done in [16]) for the first-order solution. Substituting $\sigma_{13}^{(1)}$ and $\sigma_{23}^{(1)}$ into Fourier-transformed coupled propagation equations, the output operators $a_1(L,\omega)$ and $b_1(L,\omega)$ with respect to the Fourier frequency $\omega$ can be obtained, which is a linear combination of the input operators $a_1(0,\omega)$ and $b_1(0,\omega)$ and Langevin noise terms, having the analogous expression in Ref. [16].

To test the entangled feature between the scattering field and Stokes field, as well as the atomic coherence $\sigma_{12}$, we use the criterion $V_{ij}=(\Delta u)^2+(\Delta v)^2<4$ proposed in Ref. [17], where $u=x_i\pm x_j$ and $v=p_i\mp p_j$ ("+" in $u$ and "-" in $v$ for both $V_{a1\text{-}b1}$ and $V_{\sigma_{12}\text{-}b1}$, whereas "-" in $u$ and "+" in $v$ for $V_{a1\text{-}\sigma_{12}}$) with $x_i=(a_i+a_i^+)$ and $p_i=-i(a_i-a_i^+)$. Satisfying the above inequality is sufficient to demonstrate the creation of bipartite entanglement, and the smaller the correlation V is, the stronger the degree of the bipartite entanglement becomes. As shown in Fig. 1, the Stokes field is initially in vacuum, and the scattering field is assumed to be initially in a coherent state $|\alpha\rangle$. In the following, the relevant parameters are scaled with m and MHz, or m$^{-1}$ and MHz$^{-1}$, and set according to the experimental conditions in Ref. [15] with atomic density $n_0=5\times10^{19}$, $r=1\times10^{-4}$, $L=0.06$, $\gamma_1=\gamma_2=3$, $\gamma_0=0.1$, $\omega_{12}=3036$, $\Omega_p=\Omega_c=400$, $\Delta_1=-1000$, and $\Delta_2=1000$.

Figure 2 shows the evolution of correlations $V_{a1\text{-}b1}$, $V_{a1\text{-}\sigma_{12}}$, and $V_{\sigma_{12}\text{-}b1}$ as a function of the Fourier frequency $\omega$. It can be seen that, nearly in the whole range of the Fourier frequency $\omega$, $V_{a1\text{-}b1}$, $V_{a1\text{-}\sigma_{12}}$, and $V_{\sigma_{12}\text{-}b1}$ are always less than 4, which sufficiently demonstrates that the scattering field, Stokes field, and atomic coherence are genuinely entangled with each other. In a wide range (-2600~600MHz) of the Fourier frequency $\omega$, $V_{a1-b1}$, $V_{a1-\sigma_{12}}$, and $V_{\sigma_{12}-b1}$ remain steady values of about 2, 1, and 1, except that there exist a narrow dip at the Fourier frequency $\omega$=-1000 MHz with the minimum values of about 1.11, 0.55, and 0.55, respectively, which is due to the fact that at this particular Fourier frequency, the scattering and Stokes fields are resonant with the atomic transitions 2-3 and 1-3, respectively.

The strong bipartite entanglement between the scattering field and generated Stokes field, as well as the atomic coherence can be well understood in terms of the interaction between the laser fields and atomic medium. As seen in Fig. 1a, the Stokes field $E_1$ is produced through FWM process, where every Stokes photon generation is accompanied by absorbing one scattering photon and one coupling photon and emitting one probe photon, and subsequent generation of atomic coherence excitation. Equivalently, as shown in Fig. 1b, the generated Stokes field can be regarded as the result of frequency down conversion process through mixing the scattering field with the atomic coherence prebuilt by the strong coupling and probe fields. Since a Stokes photon generation is always accompanied by annihilation of a scattering photon and creation of an atomic coherence excitation, strong tripartite entanglement can be achieved, which has the similar feature as the pump-signal-idler three-color entanglement [4]. In this respect, by using the EIT-induced atomic coherence between two different levels, a quantum frequency converter can be easily realized. The tripartite entanglement can also be seen clearly from Eqs. (4-8). Due to the atomic coherence $\sigma_{12}^{(0)}$ pre-established via EIT, the output operators $a_1(L,\omega)$ and $b_1(L,\omega)$ are both a linear combination of the input operators $a_1(0,\omega)$ and $b_1(0,\omega)$ and Langevin noise terms, that is, the output operators $a_1(L,\omega)$ and $b_1(L,\omega)$ are coupled together. If there is no atomic coherence $\sigma_{12}^{(0)}$, then the operators $a_1(L,\omega)$ and $b_1(L,\omega)$ would have no mutual coupling of $a_1(0,\omega)$ and $b_1(0,\omega)$, and no correlation would exist between the scattering and Stokes fields. Also, as seen from Eq. (6), the first-order atomic operator $\sigma_{12}^{(1)}(z,\omega)$ is a linear combination of the output operators $a_1(L,\omega)$ and $b_1(L,\omega)$ and Langevin noise terms, which implies that the atomic coherence get entangled with both the scattering field and Stokes field. In this regard, the atomic coherence acts as a quantum beam splitter, similar to the polarizing beam splitter used in the traditional way. In principle, this idea can be generalized to the case with more scattering laser fields $E_{m2}$, $E_{m3}$,….. $E_{mN}$ tuned to the vicinity of the transition $|1\rangle$-$|3\rangle$ and/or $|2\rangle$-$|3\rangle$ to mix with the same atomic coherence, then all of the scattering fields and generated Stokes/anti-Stokes fields would be entangled with the atomic coherence as well as with each other. This can be demonstrated by applying another field $E_{m2}$ scattering off the atomic coherence to examine the entangled feature among the scattering fields $E_{m1}$ and $E_{m2}$, the generated Stokes and anti-Stokes fields $E_1$ and $E_2$, and the atomic coherence excitation.

Also, under the assumptions that the detunings $\Delta_1$ and $\Delta_2$ are sufficiently large and the atomic coherence is strong enough, the evolution of the atomic operator $\sigma_{12}$ would include the terms from the interaction of the scattering field $E_{m2}$ and anti-Stokes field $E_2$ (described by operators $a_2$ and $b_2$, respectively) with the atomic medium, where the evolution of the operators $b_2$ and $a_2$ can be described by the same coupled propagation equation as $a_1$ and $b_1$. In a similar way, by Fourier transforming the Heisenberg-Langevin equations and coupled propagation equations, we can get the output operators $a_1(L,\omega)$, $a_2(L,\omega)$, $b_1(L,\omega)$, $b_2(L,\omega)$, and $\sigma_{12}(L,\omega)$.

Figure 3 shows the evolution of correlations $V_{a1\text{-}a2}$, $V_{b1\text{-}b2}$, $V_{a1\text{-}\sigma_{12}}$, and $V_{\sigma_{12}\text{-}b1}$ as a function of the Fourier frequency $\omega$. It can be seen as well that, nearly in the whole range of the Fourier frequency $\omega$, the correlations are always less than 4, which sufficiently demonstrates the generation of entanglement among the scattering fields $E_{m1}$ and $E_{m2}$, Stokes and anti-Stokes fields $E_1$ and $E_2$, as well as atomic coherence $\sigma_{12}$. The correlations $V_{a1\text{-}\sigma_{12}}$, and $V_{\sigma_{12}\text{-}b1}$ have the similar features as the case without the scattering field $E_{m2}$, which indicates the scattering field $E_{m2}$ has negligible influence on the bipartite entanglement. Also, the correlations $V_{a1-a2}$ and $V_{b1-b2}$ exhibit similar behaviors as $V_{a1-b1}$, except that there exists two narrow dips with the minimum values of about 1.58, which correspond to the two particular Fourier frequencies $\omega$=1000 MHz and $\omega$=-1000 MHz, where the scattering fields $E_{m2}$ and $E_{m1}$ are resonant with the atomic transitions 1-3 and 2-3, respectively.

Based on the above analysis, it can be seen that the EIT-induced atomic coherence can be used to conveniently perform light-light, atom-light, and atom-atom entanglement distribution. For example, two EIT-based systems A and B similar to that shown in Fig. 1a can be established by Alice and Bob, respectively. Alice use a scattering field $E_{Am1}$ mixing with the atomic coherence preproduced through EIT to generate a Stokes field $E_{A1}$, and then send the scattering field and/or Stokes field to Bob. Bob employs the scattering field $E_{Am1}$ and/or Stokes field $E_{A1}$ or another scattering field $E_{Bm1}$ mixing with the atomic coherence precreated also through EIT to generate a Stokes field $E_{B1}$ and/or an anti-Stokes field $E_{B2}$ or a Stokes (or an anti-Stokes) field $E_{B3}$. Consequently, multipartite entanglement among the scattering field $E_{Am1}$, Stokes field $E_{A1}$, Stokes and/or anti-Stokes field $E_{B1}$, $E_{B21}$, and $E_{B3}$, as well as the two atomic ensembles A and B, can be established. Therefore, by sequentially performing such operations, multipartite entanglement distribution over long distance can be easily achieved with only coherent light fields.

The robustness of entanglement is a key issue for practical quantum applications. In the present Λ-type atomic system, it critically depends in the coherence decay rate $\gamma_0$ of the lower doublet. This is clearly demonstrated in Fig. 4 by examining the influence of $\gamma_0$ on the correlations $V_{a1-b1}$, $V_{a1-\sigma_{12}}$, and $V_{\sigma_{12}-b1}$ at zero Fourier frequency. $V_{a1-b1}$, $V_{a1-\sigma_{12}}$, and $V_{\sigma_{12}-b1}$ with the initial values of about 2, 1, and 1, respectively, would increase with increasing $\gamma_0$, which means the degree of entanglement would be weakened. When $\gamma_0$ grows large enough, $V_{a1-b1}$ nearly becomes equal to 4, whereas $V_{a1-\sigma_{12}}$ and $V_{\sigma_{12}-b1}$ approximately approach 2. As is well known, in the Λ-type atomic system, the correlation time of the entangled fields and the storage time of the quantum memory are determined by the coherence decay time of the atomic lower doublet, which, in practice, can be long (~ $ms$ or even ~ $s$ [13, 18]). Therefore, this EIT-based multi-Λ-type atomic system can be a potential candidate for a quantum repeater, through which quantum state distribution over long distance can be achieved via successive entanglement exchange and distribution. In addition, as depicted in Fig. 5, bipartite entanglement between the scattering fields, Stokes and anti-Stokes fields, and the atomic coherence excitation are independent of the intensity $|\alpha|^2$ of the scattering fields. So we can use bright enough beams to carry quantum information as long as the detunings $\Delta_1$ and $\Delta_2$ of the scattering fields are sufficiently large and the atomic coherence is strong enough.

It is worth noting that the present scheme for quantum repeater is quite distinct as compared to that proposed by Yuan [13]. In Ref. [13], the Stokes or anti-Stokes field is produced through spontaneous Raman scattering and would emit in all directions, whereas in the present scheme, the generated Stokes of anti-Stokes field would propagate along a particular direction determined by the phase-matching condition for FWM [15] and the production efficiency is much higher; also, nonclassical input light fields and interferometric stabilization over long distance are required for entanglement swapping and the entanglement is created in a post-selection way, whereas in the present case, multipartite entanglement exchange and distribution can be conveniently and efficiently realized in a deterministic way via atomic coherence by using only coherent input light fields and can avoid the difficulty in keeping interferometric stabilization, which would greatly simplify the practical implementation.

In conclusion, we have proposed an efficient and convenient scheme for exchanging and distributing multipartite entangled state via atomic coherence in the multi-Λ-type atomic system. The atomic coherence induced by EIT, serves as a quantum information processor and has the function of quantum beam splitter, quantum frequency converter, quantum entangler, and quantum repeater. By using

EIT-based nondegenerate four-wave mixing processes, multipartite entangled state exchange and distribution between light and light, light and matter, as well as matter and matter can be efficiently and flexibly achieved deterministically with only coherent light fields. This method would bring great facility in realistic quantum information processing protocols, and may find potential applications in scalable quantum communication and quantum networks.

## ACKNOWLEDGEMENTS

This work is supported the National Natural Science Foundation of China (Nos. 11274225, 10974132, and 50932003), Key Basic Research Program of Shanghai Municipal Science and Technology Commission (No. 14JC1402100), and Innovation Program of Shanghai Municipal Education Commission (No. 10YZ10). Yang's e-mail is yangxih1@yahoo.com or yangxihua@yahoo.com; S.Y. Zhu's e-mail is syzhu@csrc.ac.cn.

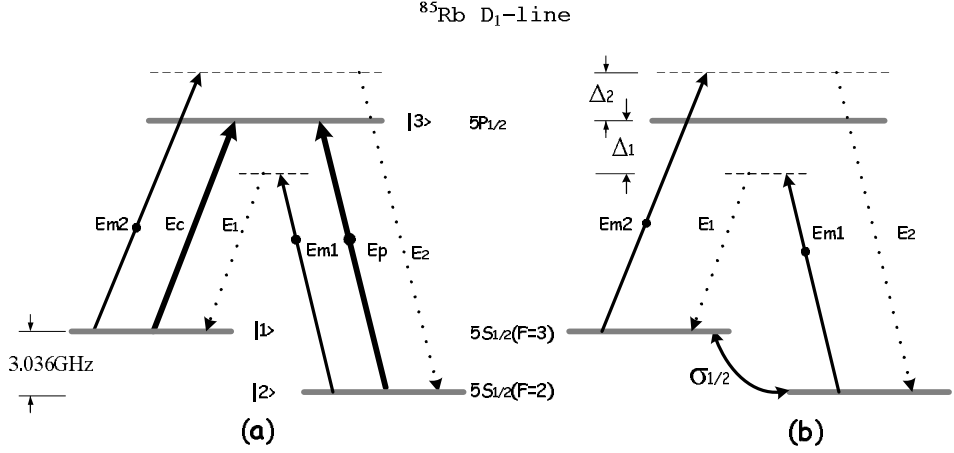

FIG. 1. (a) The triple-Λ-type system of the $D_1$ transitions in $^{85}$Rb atom coupled by the coupling ($E_c$), probe ($E_p$), and mixing ($E_{m1}$ and $E_{m2}$) fields based on the experimental configuration used in Ref. [15], where the Stokes field $E_1$ and anti-Stokes field $E_2$ are generated through two FWM processes. (b) The equivalent configuration of (a) with the lower doublet driven by the atomic coherence $\sigma_{12}$ precreated by the strong on-resonant $E_c$ and $E_p$ fields in the Λ-type EIT configuration.

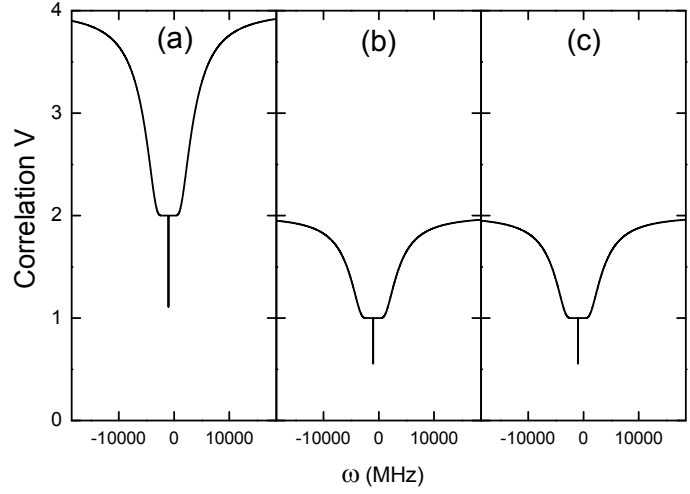

FIG. 2. The evolution of correlations $V_{a1-b1}$ (a), $V_{a1-\sigma_{12}}$ (b), and $V_{\sigma_{12}-b1}$ (c) as a function of the Fourier frequency $\omega$ with $L=0.06$, $r=1.0\times10^{-4}$, $n_0=5.0\times10^{19}$, $\gamma_1=\gamma_2=3$, $\gamma_0=0.1$, $\omega_{12}=3036$, $\Omega_p=\Omega_c=400$, and $\Delta_1=-1000$ in corresponding units of m and MHz, or m$^{-1}$ and MHz$^{-1}$.

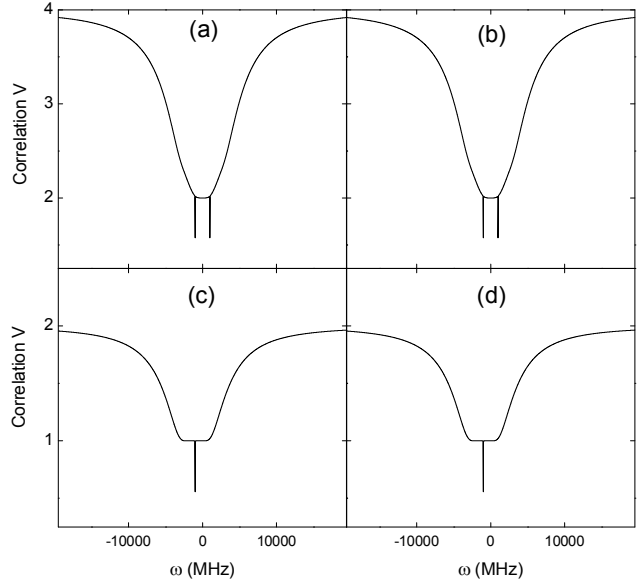

FIG. 3. The evolution of correlations $V_{a1\text{-}a2}$ (a), $V_{b1\text{-}b2}$ (b), $V_{a1\text{-}\sigma_{12}}$ (c), and $V_{\sigma_{12}\text{-}b1}$ (d) as a function of the Fourier frequency $\omega$ with $\Delta_2 = -\Delta_1 = 1000$, and the other parameters are the same as those in Fig. 2.

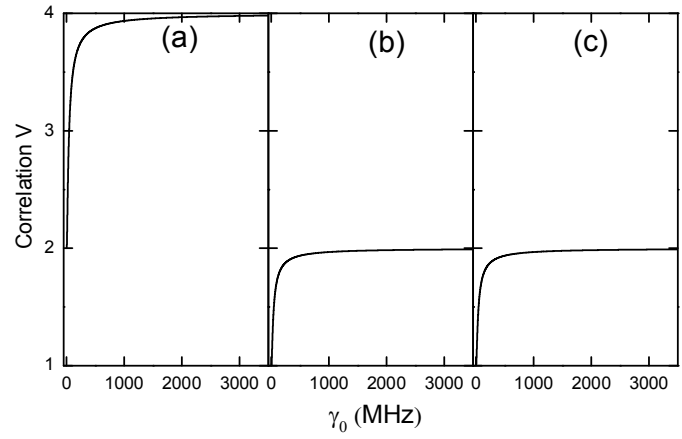

FIG. 4. The dependence of correlations $V_{a1\text{-}b1}$ (a), $V_{a1\text{-}\sigma_{12}}$ (b), and $V_{\sigma_{12}\text{-}b1}$ (c) at zero Fourier frequency on the coherence decay rate $\gamma_0$, and the other parameters are the same as those in Fig. 2.

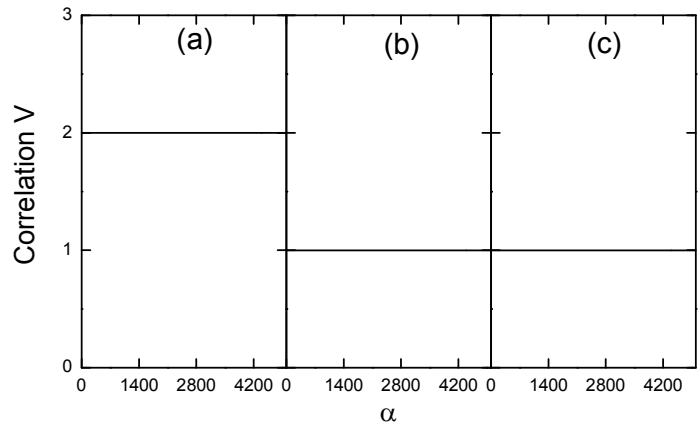

FIG. 5. The dependence of correlations $V_{a1-b1}$ (a), $V_{a1-\sigma_{12}}$ (b), and $V_{\sigma_{12}-b1}$ (c) at zero Fourier frequency on $\alpha$, and the other parameters are the same as those in Fig. 2.